\documentclass[conference]{IEEEtran}
\IEEEoverridecommandlockouts
\usepackage{hyperref}
\usepackage{cite}
\usepackage{amsmath,amssymb,amsfonts}
\usepackage{algorithmic}
\usepackage{graphicx}
\usepackage{subcaption}
\usepackage[font=small,labelfont=bf]{caption}
\usepackage{textcomp}
\usepackage{xcolor}
\usepackage{booktabs} 
\usepackage{url}
\usepackage{tikz}
\usepackage{balance}
\usetikzlibrary{positioning, shapes, arrows.meta}
\def\BibTeX{{\rm B\kern-.05em{\sc i\kern-.025em b}\kern-.08em
    T\kern-.1667em\lower.7ex\hbox{E}\kern-.125emX}}

\usepackage{tikz}
\usetikzlibrary{arrows.meta, positioning}
\tikzset{
    tick/.style = {thin},
    targ/.style = {thick},
}

\begin{document}

\title{A Deep Learning Framework for
Predicting Solar EUV Irradiance During Significant Flares} 

\author{
\IEEEauthorblockN{Sathvik Soman, Jason T. L. Wang, Haimin Wang
}
\IEEEauthorblockA{New Jersey Institute of Technology, 
Newark, NJ, USA\\
Email: \{ss5369, wangj, haimin.wang\}@njit.edu}
\and
\IEEEauthorblockN{Haodi Jiang}
\IEEEauthorblockA{Sam Houston State University, 
Huntsville, TX, USA\\
Email: haodi.jiang@shsu.edu}
}
\maketitle
\thispagestyle{plain}
\pagestyle{plain}

\begin{abstract}
We present FlareEUV, a multimodal deep learning framework for predicting daily
extreme ultraviolet (EUV) irradiance at 6.5\,nm over three consecutive days
during significant solar flares, 
using 
multi-instrument
observations from NASA's Solar Dynamics Observatory (SDO).
We consider 33 significant flares in the period 
between 2011 and 2014 in Solar Cycle 24.
The SDO observations include 13 co-aligned full-disk images,
comprising eight AIA EUV/UV and five HMI magnetic/continuum products.
FlareEUV 
learns 
the relationship between magnetic structure and coronal emission
from the raw imaging data using a lightweight attention-based 
architecture.
Our experimental results demonstrate the good performance of 
FlareEUV in 
short-term EUV irradiance forecasting during the
significant
flares and
its superiority over baseline methods.
\end{abstract}

\begin{IEEEkeywords}
Multimodal deep learning,
EUV irradiance,
Space weather
\end{IEEEkeywords}

\section{Introduction}

The Sun’s extreme ultraviolet (EUV) irradiance plays a central role in controlling the thermosphere–ionosphere energy balance, satellite drag, and radio-wave propagation. During major solar flares, EUV output can change rapidly, particularly in the 6.5\,nm band dominated by high-temperature Fe~XVIII and Fe~XX emissions \cite{aschwandencorona2019}. Accurate short-term forecasting of this wavelength is therefore essential for space–weather modeling and operational risk mitigation
\cite{tobiska2008solar}.

Traditional proxies such as F10.7 and Mg~II reproduce long-term solar variability, but fail to capture the rapid flare-driven dynamics that dominate short-term EUV evolution 
\cite{dudok2017proxies}. 
A more data-driven alternative is to identify mappings between solar observations and EUV irradiance and use them to forecast  
flare-specific
irradiance evolution. 
A major source of such observations is 
NASA’s Solar Dynamics Observatory (SDO), 
which provides continuous full-disk imaging 
from the Atmospheric Imaging Assembly (AIA) 
and vector magnetic-field measurements from 
the Helioseismic and Magnetic Imager (HMI) 
\cite{pesnell2012sdo, lemen2012aia, schou2012hmi}.  
These instruments encode rich spatial and magnetic structures—such as sunspot complexity, 
coronal loop brightness, and polarity inversion lines—which correlate strongly with 
flare-associated EUV enhancements \cite{bobra2015solarflare, aschwandencorona2019, dos2021aia}. 

In this work, we address the
following
short-term 
flare-specific
EUV irradiance forecasting problem: 
given temporally aligned multi-channel AIA and HMI observations surrounding 
a major solar flare at day $T_{0}$, the goal is to predict the 
daily
solar EUV irradiance at 6.5\,nm 
for three consecutive 
days: $T_{0}$ (flare onset/start day), 
$T_{1}$ (= $T_{0}$ + 1 day), and 
$T_{2}$ (= $T_{0}$ + 2 days).
The solar EUV irradiance at 6.5\,nm data
are extracted from
the SDO/EVE Level~3 archive available at 
the LASP Interactive Solar Irradiance Datacenter (LISIRD) 
(\url{https://lasp.colorado.edu/lisird/}).

The 6.5\,nm wavelength is dominated by 
high-temperature Fe~XVIII and Fe~XX emission lines formed at coronal temperatures 
of 6--10\,MK \cite{aschwandencorona2019}, making it a sensitive diagnostic of 
post-flare heating and magnetic energy dissipation in the upper corona. 
SDO/EVE Level~3 irradiance data provide radiometrically calibrated, daily-averaged 
full-disk measurements integrated over each UTC day, ensuring temporal consistency 
and reproducibility across events. 
These measurements serve as the reference 
standard for scientific analysis and operational space-weather forecasting, 
as they directly quantify the total solar EUV energy input to Earth's thermosphere 
and ionosphere \cite{tobiska2008solar}.

We consider significant flares during
the period between 2011 and 2014 in
Solar Cycle 24.
There are 33 flare events in the period, taken from
FlareDB 
\cite{nian2026}, whose goal
is to help scientists understand the onset of solar eruptions, such as flares and coronal mass ejections.
Each flare event in FlareDB is classified as M5.0 or greater
(including the X-class) according to NOAA Space Weather Prediction 
Center (SWPC) event reports.
 These events span a wide range of active-region morphologies and energy-release profiles, 
providing a diverse benchmark for flare-specific irradiance forecasting.
Each flare event in FlareDB is associated with
the flare start time,
peak time,
end time, as well as
  full-disk images acquired by SDO including eight AIA EUV/UV channels (94, 131, 171, 193, 211, 304, 335, 1600\,\AA) and five HMI magnetic/continuum products (B\textsubscript{field}, B\textsubscript{incl}, B\textsubscript{azim}, IC, M)
  that are in 24 hours before, and 6 hours after
  the flare event.

\subsection{Data Description}
\begin{figure}
\centering
\includegraphics[width=0.9\linewidth]{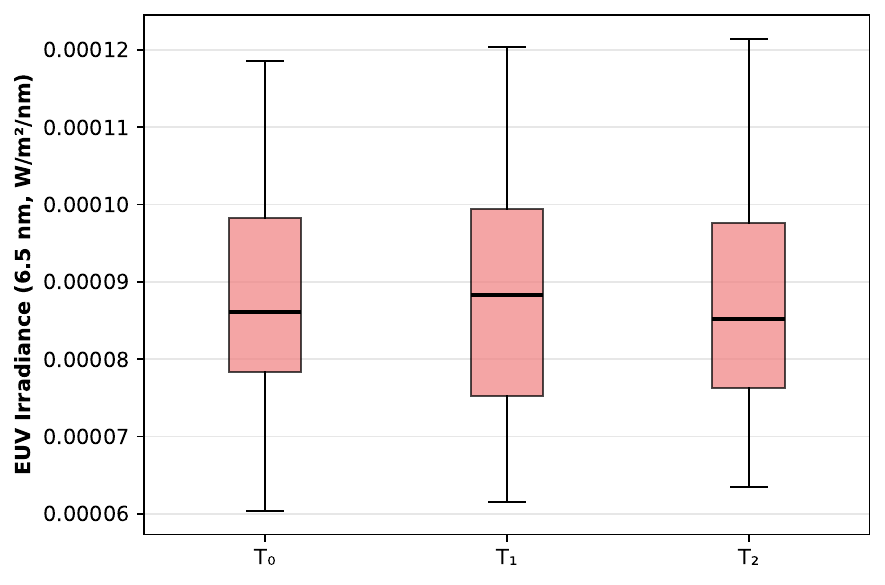}
\caption{Distribution of EVE 6.5\,nm irradiance values on three consecutive days 
$T_0$, $T_{1}$, 
and $T_{2}$ across 33 significant flares.} 
\label{fig:euv_boxplot}
\vspace{-6pt}
\end{figure}

\begin{figure}
\centering
\includegraphics[width=\linewidth]{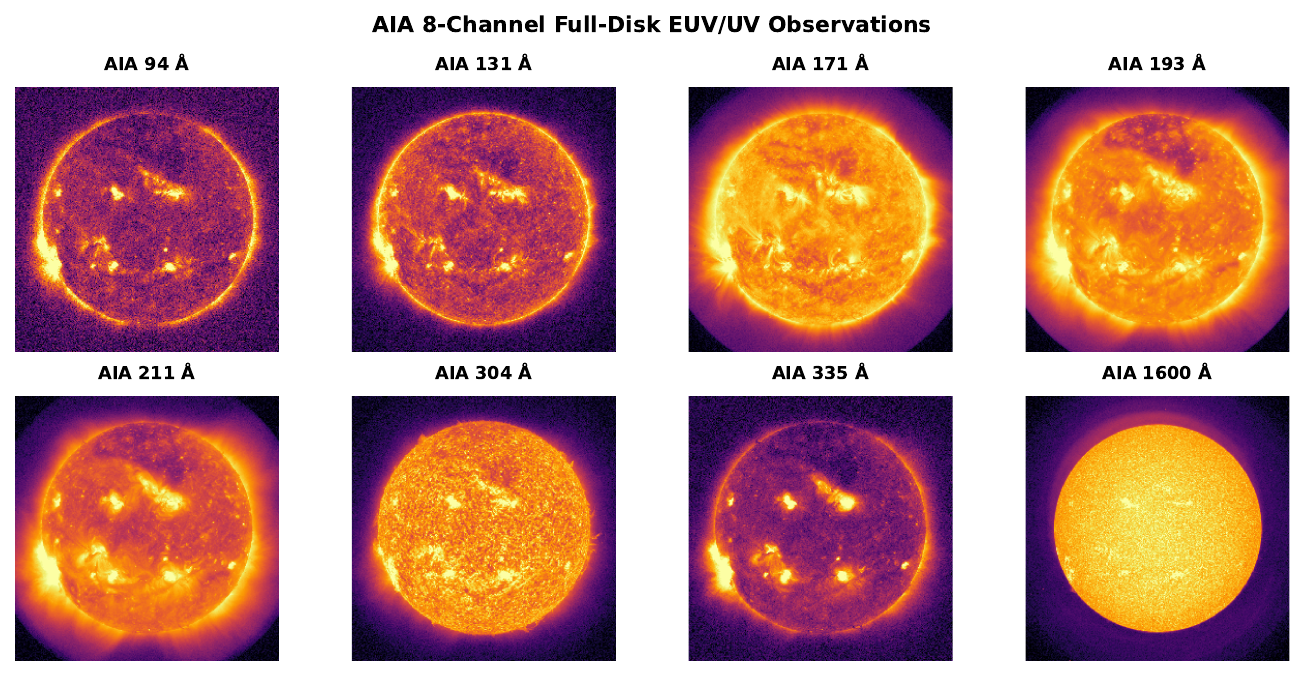}
\caption{AIA 8-channel full-disk EUV/UV observations sampled from a representative 
M6.6 flare in the active region (AR) NOAA 11158.
Each wavelength (94--1600\,\AA) captures plasma at distinct temperatures and emission lines,
providing a multi-thermal view of the corona and transition region.}
\label{fig:aia_full_disk}
\vspace{-6pt}
\end{figure}

\begin{figure}
\centering
\includegraphics[width=\linewidth]{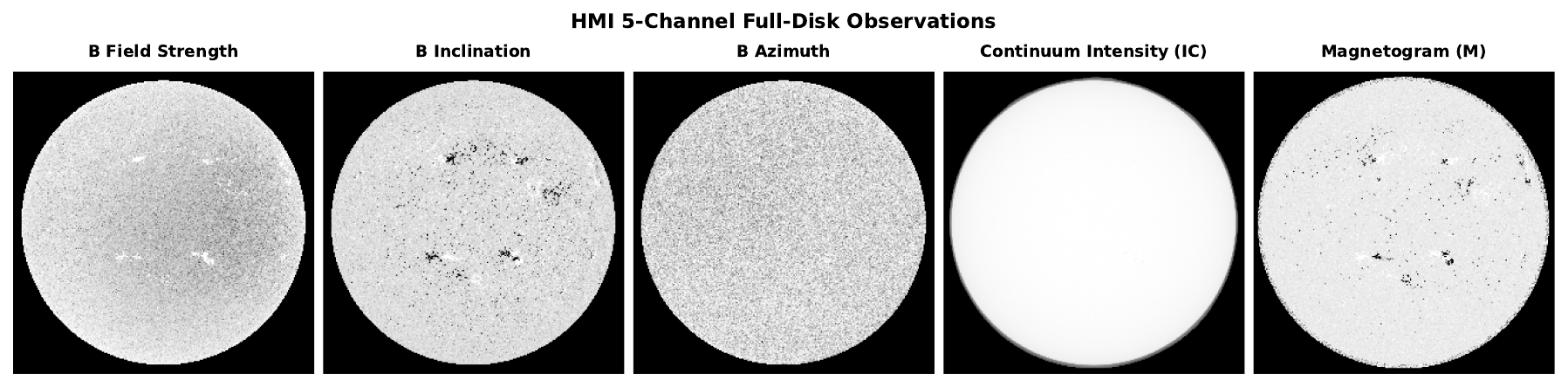}
\caption{HMI 5-channel full-disk magnetic and continuum observations for the same flare
mentioned in Fig. \ref{fig:aia_full_disk}.
Products include magnetic field strength, inclination, azimuth, continuum intensity (IC),
and line-of-sight magnetogram (M). 
}
\label{fig:hmi_full_disk}
\vspace{-6pt}
\end{figure}

Figure~\ref{fig:euv_boxplot} shows
the distribution of EVE 6.5\,nm 
irradiance targets on the three forecast days
$T_0$, $T_{1}$, $T_{2}$ 
for all 33 flare events. 
The box plots show the median (thick line), interquartile range (box), and 
full data range (whiskers) in each day.
Figures~\ref{fig:aia_full_disk} and~\ref{fig:hmi_full_disk} show preprocessed 
AIA 
and HMI images for a representative 
M6.6 flare event in the active region (AR) NOAA 11158.
AIA provides multi-wavelength EUV and UV observations tracing plasma across 
a wide temperature range, while HMI provides vector magnetic-field measurements and continuum 
structure from the photosphere. 
Together, these 13 channels form the 
harmonized 
input 
cube 
used 
for model training and testing.

\subsection{Problem Formulation}
\label{sec:formulation}

We formulate 
the problem at hand
as a supervised regression task in which pre-flare multi-instrument solar observations are mapped to post-flare irradiance evolution.
Specifically, 
at each time step, we consider a fused set of 13 co-aligned full-disk images acquired by SDO,
including
eight AIA EUV/UV channels (94, 131, 171, 193, 211, 304, 335, 1600\,\AA) and five HMI magnetic/continuum products (B\textsubscript{field}, B\textsubscript{incl}, B\textsubscript{azim}, IC, M). 
All images were preprocessed to $256{\times}256$ pixels with 
a consistent north-up orientation.

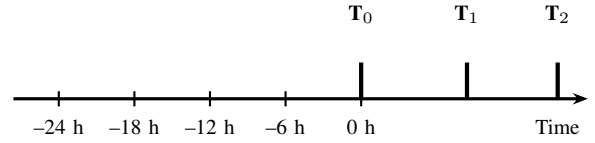
\begin{figure}
    \centering
    \begin{tikzpicture}[tick/.style={thick}, targ/.style={ultra thick}]
        \draw[very thick, -{Stealth[length=2mm]}] (0,0) -- (7.6,0);
        
        \foreach \x/\t in {0.6/--24 h, 1.6/--18 h, 2.6/--12 h, 3.6/--6 h, 
        4.6/0 h}
        {
            \draw[tick] (\x,0.10) -- (\x,-0.10);
            \node[below=1.5mm, font=\footnotesize] at (\x,0) {\t};
        }
        
        \foreach \x/\lab in {4.6/T$_0$, 6.0/T$_1$, 7.2/T$_2$} {
            \draw[targ] (\x,0) -- (\x,0.48);
            \node[above=4mm, font=\footnotesize\bfseries] at (\x,0.48) {\lab};
        }
        
        \node[below=1.5mm, font=\footnotesize] at (7.2,0) {Time};
    \end{tikzpicture}
    
    \caption{
    Formulation of our forecasting task 
    with five pre-flare inputs and three forecast outputs.
    }
    \label{fig:timeline}
    \vspace{-6pt}
\end{figure}

Let $\mathbf{X}_{t} \in \mathbb{R}^{13 \times 256 \times 256}$ denote the fused multi-instrument observation at temporal offset $t$ relative to the flare start time 
$T_{0}$.
We sample five pre-flare 
time steps (see Fig. \ref{fig:timeline}):
\[
t \in \{-24, -18, -12, -6, 0\}\ \text{hours},
\]
where $t = 0$ denotes the flare start time $T_{0}$.
The input to the FlareEUV model 
consists of the five multi-instrument observations:
\[
\mathbf{X} = [\mathbf{X}_{-24}, \mathbf{X}_{-18}, \mathbf{X}_{-12}, \mathbf{X}_{-6}, \mathbf{X}_{0}],
\]
which are temporally averaged (mean-pooled) per channel during preprocessing into a single fused tensor 
$\mathbf{X} \in \mathbb{R}^{13 \times 256 \times 256}$. 
The target output is the vector of 
daily EUV irradiance values at 6.5\,nm:
\[
\mathbf{Y} = [Y_{T_0},\, Y_{T_1},\, Y_{T_2}] \in \mathbb{R}^{3},
\]
where $Y_{T_d}$ corresponds to the 
EUV irradiance for day $d \in \{0,1,2\}$ with
$T_1 = T_0 + 24$ h and 
$T_2 = T_0 + 48$ h. 
Thus, the forecasting task at hand is defined as:
\[
f_{\theta}: \mathbb{R}^{13 \times 256 \times 256} \rightarrow \mathbb{R}^{3},
\]
where $f_{\theta}$ is a learnable deep neural network, trained on the major flares considered in the study.

\subsection{Contributions of the Work}

Our work makes the following  
methodological and
experimental advances for flare-specific EUV irradiance 
forecasting:

1. Unified 13-channel multimodal fusion.
    To solve the problem at hand, we propose FlareEUV, a lightweight multimodal deep learning framework that integrates
    EUV/UV radiative structure (AIA) and photospheric magnetic topology (HMI) via a widened
    13-channel input stem. The backbone is based on ResNet-34 with
    squeeze-and-excitation blocks for channel-wise attention
    \cite{he2016resnet, hu2018senet},
    providing improved feature recalibration over vanilla residual networks.
    ImageNet-pretrained initialization \cite{deng2009imagenet} enables stable learning on limited flare-level samples. In addition,
    we design a horizon-weighted loss that prioritizes accurate forecasting at $T_0$ while still supervising $T_1$ and $T_2$, mirroring the natural decay of predictive confidence across the three-day horizon.
     
   2. Cross-flare evaluation and multimodal ablation.
    We analyze 33 significant flare events 
    during
the period between 2011 and 2014 in
Solar Cycle 24 
    using a strict leave-one-flare-out (LOFO) protocol,
    in which each fold has 31 training events, 1 validation event, and 1 test event, and there are 33 folds in total. 
    With the LOFO protocol, 
    AIA+HMI fusion consistently outperforms AIA-only and HMI-only variants, demonstrating the complementary roles of radiative and magnetic information.

There are two studies that are closely related to our work.
Szenicer et al. (2019) \cite{szenicer2019deep} 
developed a convolutional neural network to 
map AIA images to EUV spectral irradiance measurements.
Jiang et al. (2025)
\cite{2025ApJS..280...50J}
reconstructed EUV irradiance from Ca II K images
with Bayesian deep learning.
In contrast to the above studies, our work focuses on
predicting EUV irradiance during significant flares
using
both AIA and HMI images.

The remainder of this paper is organized as follows.
Section \ref{sec:model} presents the FlareEUV model
and details its architecture.
Section \ref{sec:results} 
reports our experimental results.
Section \ref{sec:conclusion}
concludes the paper.

\section{The FlareEUV Model}
\label{sec:model}

\subsection{Motivation and Design Philosophy}

Forecasting EUV irradiance from full-disk solar imagery requires modeling the nonlinear coupling between photospheric magnetic fields and coronal radiative output.  
AIA channels encode multi-thermal coronal emission spanning 0.6--20\,MK, while HMI products describe the underlying photospheric magnetic topology that governs magnetic energy storage, flare heating, and post-flare coronal evolution.  
Recent deep-learning studies have demonstrated the feasibility of learning these magnetic--radiative relationships directly from solar imagery or magnetograms \cite{park2019generation, szenicer2019deep}. 
An effective model must therefore fuse these complementary modalities while preserving full-disk spatial context and remaining robust to strong inter-event variability across diverse active region configurations.
Moreover, traditional convolutional backbones pretrained on natural images (e.g., ResNet, VGG) provide strong spatial feature extractors that transfer well to heliophysics imagery \cite{nigam2018deepflare}, but require architectural adaptation to handle the 13-channel multimodal input tensor combining AIA and HMI observations.

With the above considerations in mind, FlareEUV adopts 
a hybrid design strategy:
\begin{enumerate} 
    \item Compact architecture: ResNet-34 
    \cite{he2016resnet} is employed, which provides sufficient representational capacity for full-disk $256{\times}256$ imagery while remaining trainable under leave-one-flare-out (LOFO) cross-validation with 31 training events per fold.
    \item Transfer learning from ImageNet: Initializing convolutional layers with weights pretrained on natural images stabilizes optimization and reduces event complexity, enabling effective learning from limited heliophysics data \cite{park2019generation}.
    \item Multimodal input stem: The first convolutional layer is expanded from 3 channels (for RGB) to 13 channels to jointly process AIA and HMI observations, with new channel weights initialized as the mean of pretrained RGB filters to preserve learned low-level features (edges, textures).
    \item Channel attention mechanisms: 
    Squeeze-and-excitation (SE) blocks \cite{hu2018senet} are inserted after each residual stage to dynamically reweight channel-wise features, allowing the model to emphasize the most predictive wavelengths and magnetic field components on a per-event basis.
\end{enumerate}

This design yields a compact yet expressive architecture capable of extracting cross-flare magnetic--radiative patterns critical for short-horizon EUV forecasting, balancing model capacity with the limited availability of major flare events.

\subsection{Model Architecture}
\begin{figure*}
[t]
\centering
\includegraphics[width=1\linewidth]{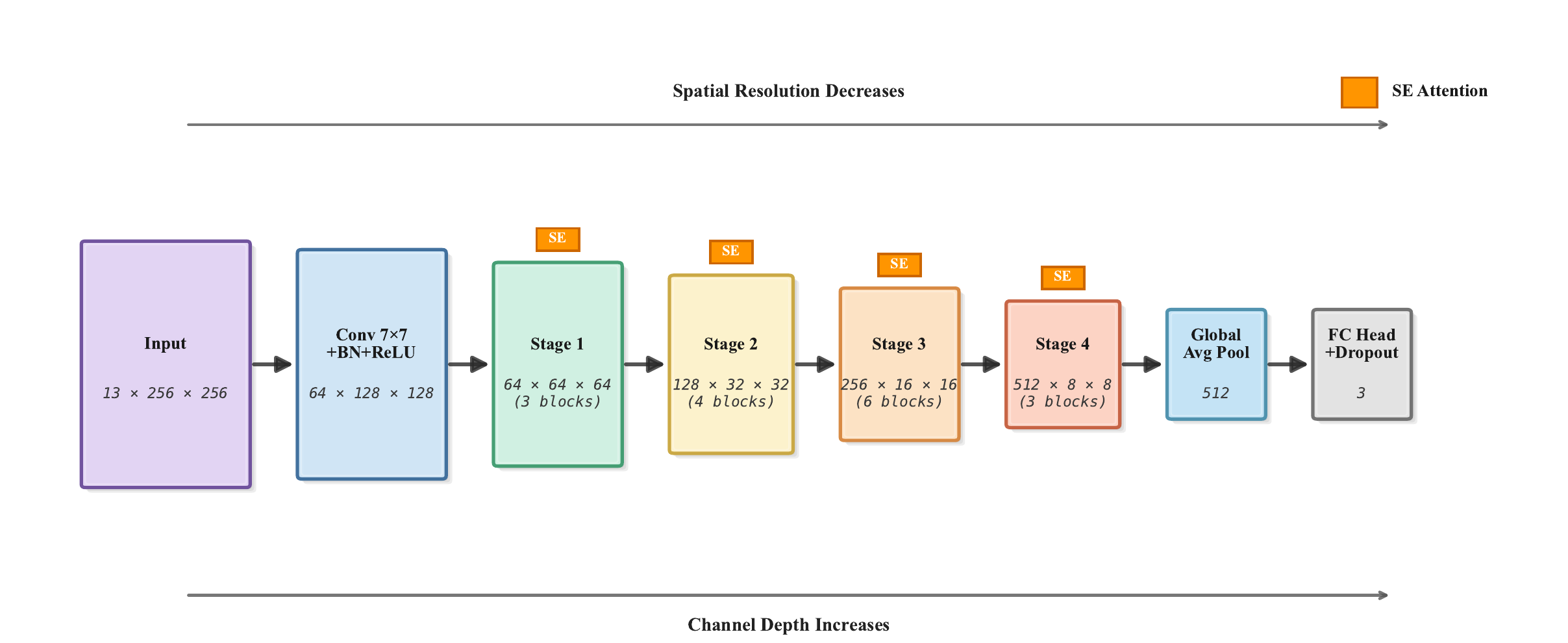}
\caption{FlareEUV architecture based on 
ResNet-34 with squeeze-and-excitation attention. The 13-channel input (8 AIA + 5 HMI) passes through a 7×7 convolutional stem followed by four residual stages (3, 4, 6, 3 blocks respectively) with progressive spatial downsampling (256$\to$8 pixels) and channel expansion (13$\to$512). Orange boxes indicate 
squeeze-and-excitation (SE) attention blocks inserted after each residual stage for dynamic channel reweighting. Global average pooling (GAP) aggregates spatial features before the two-layer regression head that outputs three forecast values at $T_0$ (flare onset), $T_{1}$ (24\,h ahead), 
and $T_{2}$ (48\,h ahead),
respectively.}
\label{fig:architecture}
\end{figure*}

FlareEUV is an enhanced ResNet-34, augmented with channel attention mechanisms to handle multimodal solar imagery.
Figure~\ref{fig:architecture} presents the
architecture of FlareEUV, which consists of:
\begin{enumerate} 
    \item Input stem: A $7{\times}7$ convolutional layer with stride 2, accepting $13{\times}256{\times}256$ input tensors with 8 AIA plus 5 HMI channels, followed by batch normalization, ReLU activation and $3{\times}3$ maximum pooling with stride 2.
    \item Residual stages: Four stages of residual blocks (3, 4, 6, 3 blocks, respectively) with channel dimensions [64, 128, 256, 512] and progressive spatial downsampling via strided convolutions.
    \item Channel attention: Squeeze-and-excitation blocks (reduction ratio $r=16$) inserted after each residual stage to dynamically reweight feature channels based on the global spatial context.
    \item Regression head: Global average pooling followed by a two-layer fully-connected network with dropout (0.3, 0.15) mapping 512-dimensional feature vectors to 3 output values 
    (irradiance values at $T_0$, $T_{1}$, $T_{2}$, respectively).
\end{enumerate}

\subsection{Training Configuration and Protocol}

All input tensors were normalized per-fold using channel-wise means and standard deviations computed exclusively from the 31 training flares in each leave-one-flare-out fold. 
This per-fold normalization strategy prevents data leakage from validation and test flares, ensuring that normalization parameters reflect only historical observations available at training time.
The EVE irradiance targets were zero-mean normalized using a fixed scale factor of $10^{-4}$ to stabilize regression on small-magnitude values
(see Fig. \ref{fig:euv_boxplot}).

Training was performed using the AdamW optimizer \cite{loshchilov2017adamw} with learning rate $\eta = 3{\times}10^{-4}$, weight decay $\lambda = 10^{-4}$, momentum parameters $\beta = (0.9, 0.999)$, and batch size 2.
To stabilize training on limited 31-event folds, several regularization 
strategies were employed. 
A cosine annealing scheduler with warm restarts 
periodically 
resets the learning rate to $\eta = 3{\times}10^{-4}$, allowing the optimizer 
to escape local minima. The restart period doubles after each cycle (20, 40, 
80 epochs), enabling rapid exploration followed by finer convergence—beneficial 
for small datasets with spurious loss surface attractors.
Gradient clipping (max L2 norm = 1.0) prevents exploding gradients during 
backpropagation through 34 layers. With batch size = 2, gradients can increase 
by orders of magnitude on outlier flares, and clipping ensures bounded 
parameter updates for stable convergence.
Early stopping with 50-epoch patience prevented overfitting. If validation 
loss failed to improve for 50 consecutive epochs, training terminated and 
the best checkpoint was restored. Training ran up to 100 epochs per fold, 
typically stopping at 50-80 epochs.

Data augmentation via random horizontal and vertical flips (50\% probability) 
artificially expanded the training set
in each fold. 
These transformations preserve physical 
content while introducing spatial diversity (east-west and north-south 
hemispheric symmetry). 
No augmentation was applied during validation or 
testing to ensure that the metrics reflect the true generalization. Rotation-based 
augmentation was excluded to preserve the north-up coordinate convention.

\subsection{Loss Function}

To prioritize accurate prediction of 
EUV irradiance while still supervising longer horizons, the model minimizes a horizon-weighted mean squared error.
Let $Y^{(i)}_{T_d}$ and $\hat{Y}^{(i)}_{T_d}$ denote the true and predicted irradiance values for event $i$ at the forecast horizon $d \in \{0,1,2\}$ 
(corresponding to $T_0$, $T_{1}$, and 
$T_{2}$).
The loss is defined as:
\[
\mathcal{L} = \frac{1}{N} \sum_{i=1}^{N} \sum_{d=0}^{2} \mathbf{w}_d \bigl( Y^{(i)}_{T_d} - 
\hat{Y}^{(i)}_{T_d} \bigr)^2,
\qquad
\mathbf{w} = [0.5,\, 0.3,\, 0.2],
\]
where $N$ is the batch size and $\mathbf{w}_d$ are horizon-specific weights.
The weights $\mathbf{w} = [0.5, 0.3, 0.2]$ enforce higher precision at the flare onset time ($T_0$) with diminishing emphasis on $T_{1}$ (24\,h ahead) and $T_{2}$ (48\,h ahead), reflecting both the operational priority of near-term forecasts and the physical decay of predictive skill with increasing temporal distance from $T_0$.
This horizon-aware loss stabilizes training,  
preventing dominance by outlier flares with extreme irradiance values.

\subsection{Model Effectiveness and Efficiency}

As described above, FlareEUV employs a convolutional architecture with
residual blocks and channel attention mechanisms.
Despite 34 layers,
FlareEUV remains computationally 
efficient. The 13-channel input stem adds only 11.3M parameters,
necessary for multimodal AIA+HMI input. Training a single 
LOFO fold (31 flares, up to 100 epochs) requires 45--60 minutes on 
NVIDIA GPU with less than 8\,GB memory. Inference completes in 10--20\,ms per flare, 
enabling real-time forecasting. This balance between expressiveness, 
interpretability, and efficiency makes FlareEUV suitable for both scientific 
analysis and operational space-weather systems.

\section{Experiments and Results}
\label{sec:results}

\subsection{Evaluation Metrics}
\label{sec:metrics}

We assess predictive performance using two complementary metrics,
defined below,
computed on aggregated predictions across all 33 folds and all three forecast horizons, totally 99 prediction–target pairs with 33 test flares $\times$ 3 horizons.
Let $Y_{i,d}$ and $\hat{Y}_{i,d}$ denote the true and predicted irradiance values for the test flare $i$ on the horizon $d\in\{0,1,2\}$.

Pearson Correlation Coefficient 
(PCC):
This metric
quantifies the linear association between predicted and observed irradiance values:
\begin{equation}
\text{PCC} = \frac{\sum_{i,d} (\hat{Y}_{i,d} - \bar{\hat{Y}})(Y_{i,d} - \bar{Y})}
         {\sqrt{\sum_{i,d} (\hat{Y}_{i,d} - \bar{\hat{Y}})^2}
          \,\sqrt{\sum_{i,d} (Y_{i,d} - \bar{Y})^2}},
\end{equation}
where $\bar{Y}$ and $\bar{\hat{Y}}$ denote the global means of observed and predicted irradiance values.
Statistical significance is assessed using the corresponding $p$-value under the null hypothesis of zero correlation.

Mean Absolute Error (MAE):
This metric
measures the mean prediction error in physical units (W/m$^2$/nm):
\begin{equation}
\text{MAE} = \frac{1}{3N} \sum_{i=1}^{N} \sum_{d=0}^{2} \lvert Y_{i,d} - \hat{Y}_{i,d} \rvert,
\end{equation}
where $N=33$ is the total number of test flares, and $d \in \{0,1,2\}$ indexes the three forecast horizons.
In addition, 
we use PCC and MAE at
$T_{0}$,
$T_{1}$,
$T_{2}$,
respectively,
to represent per-horizon metrics.
These 
per-horizon metrics  
are computed to evaluate whether predictive skill degrades with increasing temporal distance from 
the flare onset time $T_{0}$.

\subsection{Ablation Study}

To quantify the contribution of radiative (EUV) and magnetic information (HMI), we compare three input configurations derived from the Solar Dynamics Observatory (SDO):
\begin{itemize}
    \item HMI-only: 5 magnetic and continuum products (B\textsubscript{field}, B\textsubscript{incl}, B\textsubscript{azim}, IC, M) encoding photospheric magnetic structure,
    \item AIA-only: 8 EUV/UV channels (94--1600\,\AA) encoding multi-thermal coronal emissions,
    \item AIA+HMI: 13-channel fused representation combining both radiative (AIA) and magnetic structure (HMI).
\end{itemize}

\subsubsection{Global Predictive Performance}
Table~\ref{tab:ablation_global} summarizes the global performance of each modality across all 33 flares and three forecast horizons,
ending with 99 prediction–target pairs,
under LOFO cross-validation.
The best metric values are highlighted in boldface.
The multimodal AIA+HMI FlareEUV fusion model achieves the strongest predictive skill, with PCC = 0.927 ($p$ $<10^{-40}$) and MAE =  $4.41 \times 10^{-6}$~W/m$^2$/nm.  
Compared to the single-modality baselines, fusion reduces the error by 18.0\% relative to AIA-only and 39.7\% relative to HMI-only.

\begin{table}[htbp]
\centering
\caption{Performance Metric Values under LOFO Cross-Validation in the Ablation Study}
\label{tab:ablation_global}
\begin{tabular}{lcc}
\hline
\textbf{Model} & PCC & MAE ($\times10^{-6}$ W/m$^{2}$/nm) \\
\hline
HMI-only      & 0.835 & 7.32 \\
AIA-only      & 0.868 & 5.38 \\
AIA+HMI     & \textbf{0.927} & \textbf{4.41} \\
\hline
\end{tabular}
\end{table}

\begin{figure}
\centering
\begin{subfigure}{\linewidth}
    \centering\includegraphics[width=\linewidth,height=0.25\textheight,keepaspectratio]{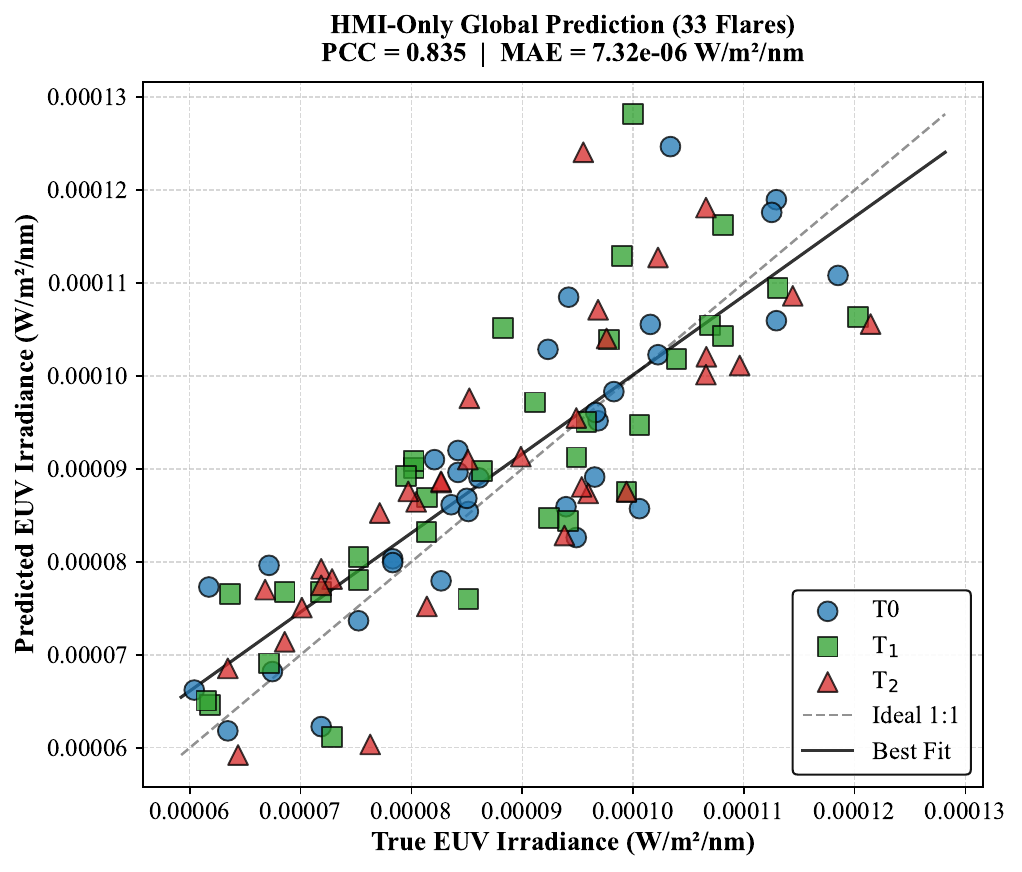}
    \caption{HMI-only}
    \vspace*{+3mm}
\end{subfigure}
\begin{subfigure}{\linewidth}
    \centering\includegraphics[width=\linewidth,height=0.25\textheight,keepaspectratio]{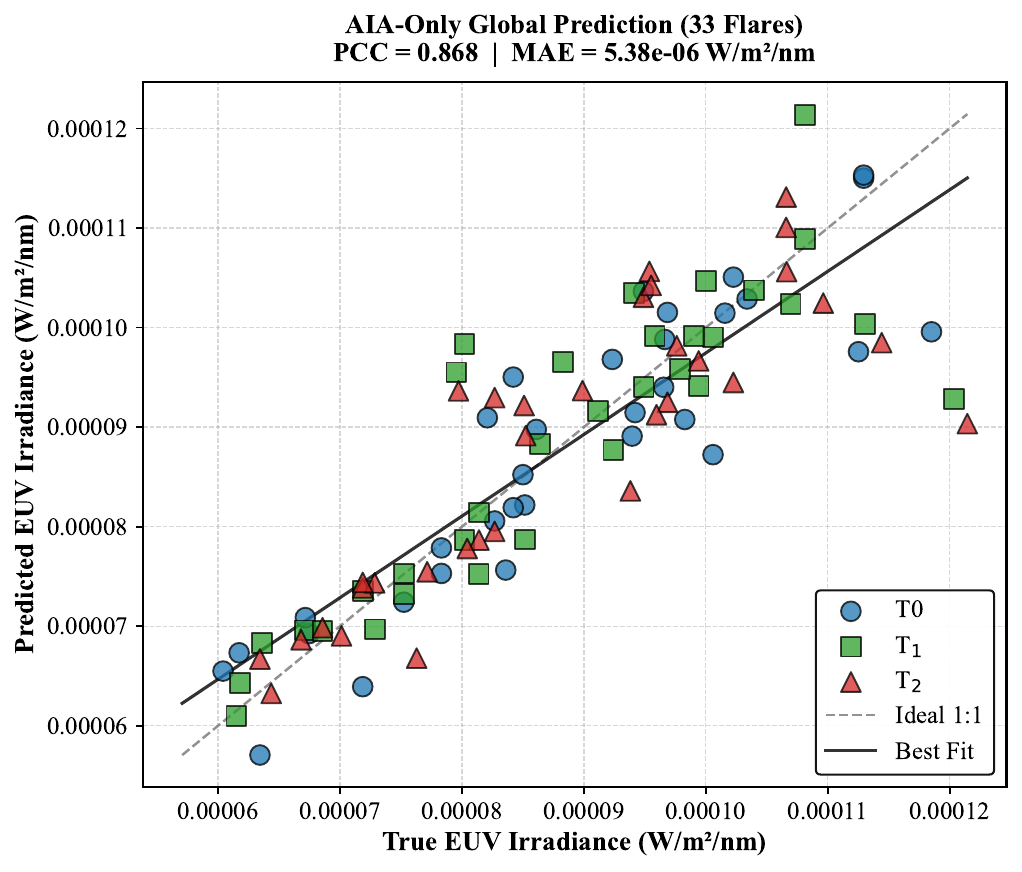}
    \caption{AIA-only}
    \vspace*{+3mm}
\end{subfigure}
\begin{subfigure}{\linewidth}
    \centering  \includegraphics[width=\linewidth,height=0.25\textheight,keepaspectratio]{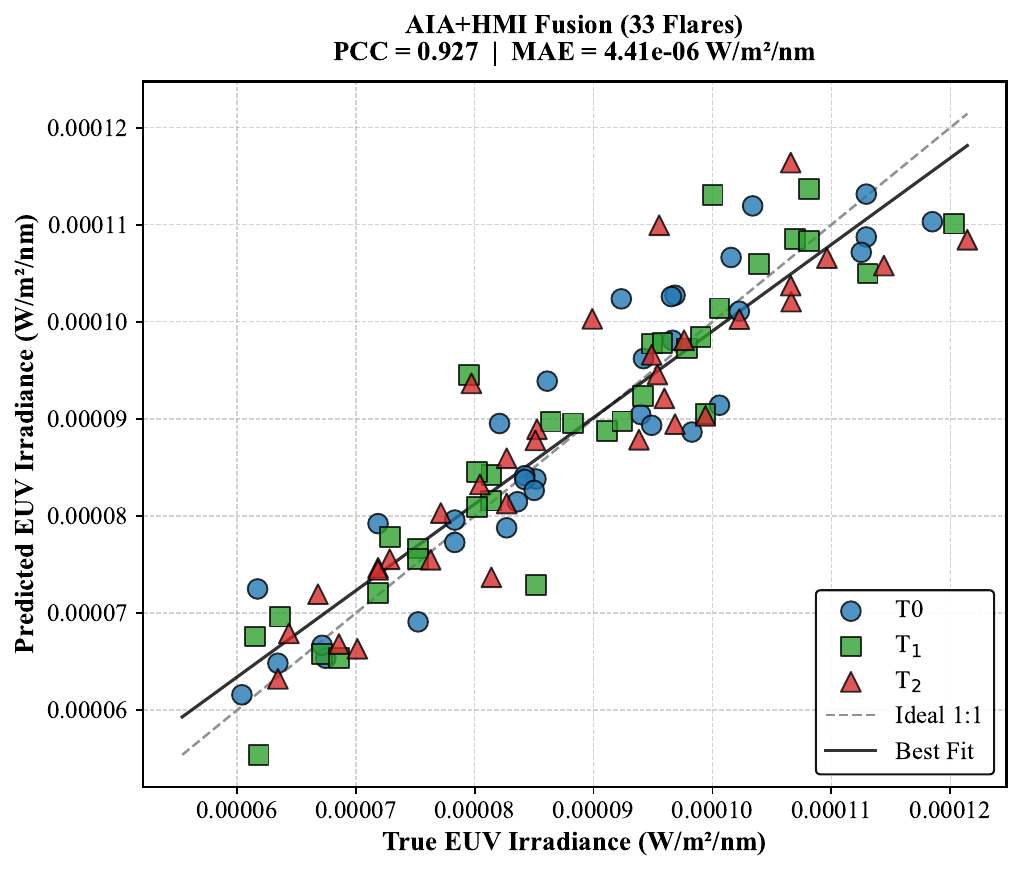}
    \caption{AIA+HMI (Fusion)}
        \vspace*{+3mm}
\end{subfigure}
\caption{
Global predictive performance under modality ablation.
All subfigures share identical axes for fair comparison.
The AIA+HMI fusion model exhibits the strongest clustering along the ideal 1:1 line, demonstrating superior predictive consistency and long-horizon generalization.
}
\label{fig:scatter_global_comparison}
\end{figure}

This modality hierarchy—HMI-only 
$<$ AIA-only 
$<$ AIA+HMI—is physically consistent.
AIA channels provide direct radiative signatures of coronal plasma (more strongly correlated with EUV irradiance), while HMI magnetic context encodes the underlying energy reservoir in the photosphere that drives post-flare EUV evolution.
The improvement from fusion demonstrates that magnetic topology provides complementary physical information that is not captured in instantaneous AIA emission patterns alone.
Figure~\ref{fig:scatter_global_comparison}
further shows that the fused model achieves a visibly tighter clustering around the ideal 1:1 line, particularly at $T_{2}$, indicating a stronger long-term generalization.

\subsubsection{Horizon-wise Performance Analysis}

Table~\ref{tab:ablation_horizonwise} summarizes the per-horizon predictive performance across all model configurations.
The fused AIA+HMI model achieves the strongest results at all three horizons, consistently achieving the highest Pearson correlation and the lowest MAE.
Notably, the benefit of multimodal fusion increases with a longer forecast horizon: 
at $T_0$ (flare onset), the improvement over 
AIA-only is approximately +3.5\%, 
while at $T_{2}$ (48\,h ahead), the gain reaches +8.7\%.

This behavior is physically meaningful and reflects the distinct temporal 
evolution of photospheric magnetic fields versus coronal radiative emission. 
Photospheric magnetic fields (HMI) evolve through flux emergence, shear 
buildup along polarity inversion lines, and gradual active-region restructuring 
over hour-to-day timescales driven by subsurface convection and magnetic 
buoyancy. Once an active region is established, its large-scale magnetic 
configuration (total flux, field strength distribution, polarity separation) 
remains relatively stable over 24--48\,hours despite local changes, providing 
persistent information about the energy reservoir available for future flares. 
Thus, pre-flare HMI observations encode stable longer-term predictive cues 
that remain informative for multi-day forecasting, as photospheric topology 
governs the coronal heating budget
\cite{2025MNRAS.539.1820H}.

In contrast, coronal radiative signatures (AIA) reflect instantaneous plasma 
conditions (temperature, density, emission measure) that fluctuate rapidly 
(minutes to hours) in response to transient heating, magnetic reconnection, 
and wave-driven energy transport. AIA provides excellent snapshots of the current 
coronal state and correlates directly with immediate flare-day irradiance 
($T_0$), but pre-flare AIA observations become progressively less reliable 
for 24--48\,h forecasting as the corona continuously responds to evolving 
photospheric stress. The superior performance of AIA+HMI fusion at longer 
horizons demonstrates that integrating stable magnetic context with 
instantaneous radiative information enables more accurate post-flare 
extrapolation than coronal emission alone. This aligns with established 
solar physics: photospheric fields set boundary conditions for coronal 
energy storage, whereas coronal emission reflects instantaneous heating and 
cooling processes.
These results demonstrate that magnetic context plays an increasingly dominant role in multi-day irradiance prediction, and that its integration is essential for robust operational space-weather forecasting.

\begin{table}
\centering
\caption{Per-Horizon Predictive Performance under LOFO Cross-Validation
in the Ablation Study} 
\label{tab:ablation_horizonwise}
\begin{tabular}{lccc|ccc}
\hline
& \multicolumn{3}{c}{\textbf{PCC}} & \multicolumn{3}{c}{\textbf{MAE ($\times10^{-6}$)}} \\
\cmidrule(lr){2-4} \cmidrule(lr){5-7}
\textbf{Model} & \textbf{$T_0$} & \textbf{$T_{1}$} & \textbf{$T_{2}$} & \textbf{$T_0$} & 
\textbf{$T_{1}$} & \textbf{$T_{2}$} \\
\hline
HMI-only      & 0.867 & 0.830 & 0.809 & 6.25 & 7.52 & 8.18 \\
AIA-only      & 0.903 & 0.862 & 0.840 & 5.11 & 5.09 & 5.94 \\
AIA+HMI & \textbf{0.935} & \textbf{0.932} & \textbf{0.913} & \textbf{4.34} & \textbf{4.04} & \textbf{4.85} \\
\hline
\end{tabular}
\vspace{-7pt}
\end{table}

A notable trend is that both AIA-only and AIA+HMI models achieve their lowest MAE at $T_{1}$ (24\,h ahead), rather than at flare onset.
This behavior is consistent with the typical evolution of post-flare EUV emission, which often reaches its thermal peak within the first 24 hours before gradually decaying
\cite{2023A&A...675A.147C}.
The relatively small increase in error at $T_{2}$ further suggests that the fused model captures stable magnetic–radiative dependencies that remain informative beyond the immediate flare phase, enabling reliable multi-day generalization across diverse solar events.

\subsection{Baseline Performance Comparison}

To rigorously evaluate the contribution of architectural design choices, we compare FlareEUV with two baseline models under identical training conditions, data splits, and evaluation protocols:

\begin{itemize}
    \item Baseline CNN: A lightweight 5-layer convolutional network with batch normalization and dropout, trained from random initialization (1.72M parameters).
    \item Vanilla ResNet-34: Standard ResNet-34 with ImageNet pretraining and 13-channel input adaptation, but without 
    squeeze-and-excitation (SE) attention blocks (21.45M parameters).
    \item FlareEUV: ResNet-34 with ImageNet pretraining, 13-channel input, and four SE attention blocks inserted after residual stages (21.49M parameters).
\end{itemize}

\begin{figure}
    \centering
    \includegraphics[width=\columnwidth]{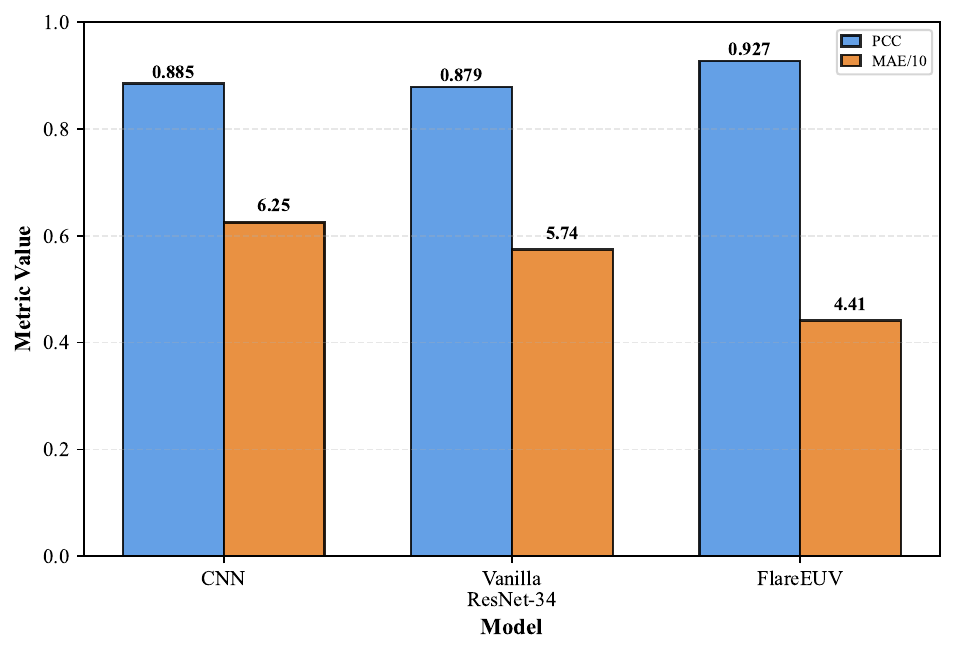}
    \caption{Predictive performance comparison among Baseline CNN, Vanilla ResNet-34, and FlareEUV models in terms of PCC and MAE.
    Each metric value in the figure represents the average value across 
    forecast horizons
    $T_{0}$, $T_{1}$, and $T_{2}$.}
    \label{fig:fig7}
\end{figure}

All models 
used the same 13-channel AIA+HMI input and training protocol.
Specifically, they
were trained using identical hyperparameters 
(AdamW optimizer, learning rate = $3{\times}10^{-4}$, cosine annealing scheduler, 
batch size = 2) and evaluated under the same LOFO cross-validation protocol across 33 major flares and three forecast horizons. 
Table~\ref{tab:baseline_comparison} 
compares the number of parameters used by the models.
Figure \ref{fig:fig7} summarizes
the results obtained by comparing the models.
It can be seen from Table~\ref{tab:baseline_comparison} 
and Figure \ref{fig:fig7}
that
SE attention adds only 43.5K parameters (0.2\% overhead
vs. Vanilla ResNet-34)
but provides substantial performance gains.

\begin{table}[h]
\centering
\caption{Number of Parameters Used by 
Three Models} 
\label{tab:baseline_comparison}
\vspace{2mm}
\begin{tabular}{lc}
\hline
\textbf{Model} & \textbf{Parameters} 
\\
\hline
Baseline CNN          & 1.72M 
\\
Vanilla ResNet-34     & 21.45M 
\\
FlareEUV & 21.49M 
\\
\hline
\end{tabular}
\end{table}

\noindent\textbf{Why SE attention helps.}
Squeeze-and-excitation (SE) blocks improve multimodal fusion by adaptively 
reweighting the 13 input channels for each flare. The SE mechanism operates 
in two stages: (1) the global average pooling aggregates each feature map into a 
channel descriptor $z_c$ that captures the global response magnitude; (2) a compact 
excitation network learns attention weights:
\[
\alpha_{c} = \sigma(W_{2} \, \text{ReLU}(W_{1} \, z_{c})),
\]
where $W_1, W_2$ are learnable matrices with
the reduction ratio $r=16$, and 
$\alpha_{c} \in [0,1]$ is applied by element-wise multiplication. The sigmoid function $\sigma$
ensures bounded weights while ReLU enables modeling complex inter-channel 
dependencies (e.g., AIA 94\,\AA{} hot emission with HMI strong fields).

This adaptive reweighting is critical because different flares require different 
channel combinations
\cite{2008LRSP....5....1B,
2011ApJ...727L..52R,
2012A&A...540A..24B}.
For example,
X-class solar flares derive a significant signal from the AIA 94 \AA{} and 131 \AA{} channels because these wavelengths are specifically designed to capture the emission from extremely hot plasma 
(6--10\, MK)
produced during major flaring events.
However, 
for M-class solar flares, the AIA 193 \AA{} and 211 \AA{} channels are critical for the evolution of post-flare loop systems at temperatures between 1--2\, MK. 
Vanilla ResNet-34 treats all channels equally, 
forcing a single fixed weighting that is suboptimal for various morphologies. 
SE blocks 
learn event-specific importance, asking: 
"for this active region, which AIA/HMI 
combination predicts its EUV evolution?"

Adding SE attention blocks to Vanilla ResNet-34 (+43.5K parameters, +0.2\% overhead) 
to obtain FlareEUV
yields a +5.5\% correlation 
improvement and a 23.2\% MAE reduction
compared to the Vanilla ResNet-34 baseline, 
demonstrating that lightweight attention mechanisms can unlock substantial performance gains from deep architectures without significant computational overhead. 
This result shows that
the FlareEUV model learned physically 
meaningful channel combinations—hot emission + strong fields for impulsive events, 
ambient structure + shear for gradual events—rather than simply adding capacity.

Figure~\ref{fig:fig7}  
provides a comprehensive visual comparison of the three models in multiple evaluation dimensions. 
It can be seen in Figure~\ref{fig:fig7} 
that FlareEUV achieves superior performance in both correlation (PCC = 0.927) and absolute error 
(MAE = $4.41{\times}10^{-6}$~W/m$^2$/nm), substantially outperforming both baselines.
We note that compared to the best-performing baseline (Baseline CNN, PCC = 0.885), FlareEUV improves the correlation by 4.7\% and reduces MAE by 29.4\% (from $6.25{\times}10^{-6}$ to $4.41{\times}10^{-6}$~W/m$^2$/nm). 
The Baseline CNN, despite having only 1.72M parameters (8\% of ResNet-34's capacity), achieves competitive performance, demonstrating that effective feature extraction from full-disk solar imagery does not strictly require very deep architectures. 

\begin{figure}
\centering
\includegraphics[width=\columnwidth]{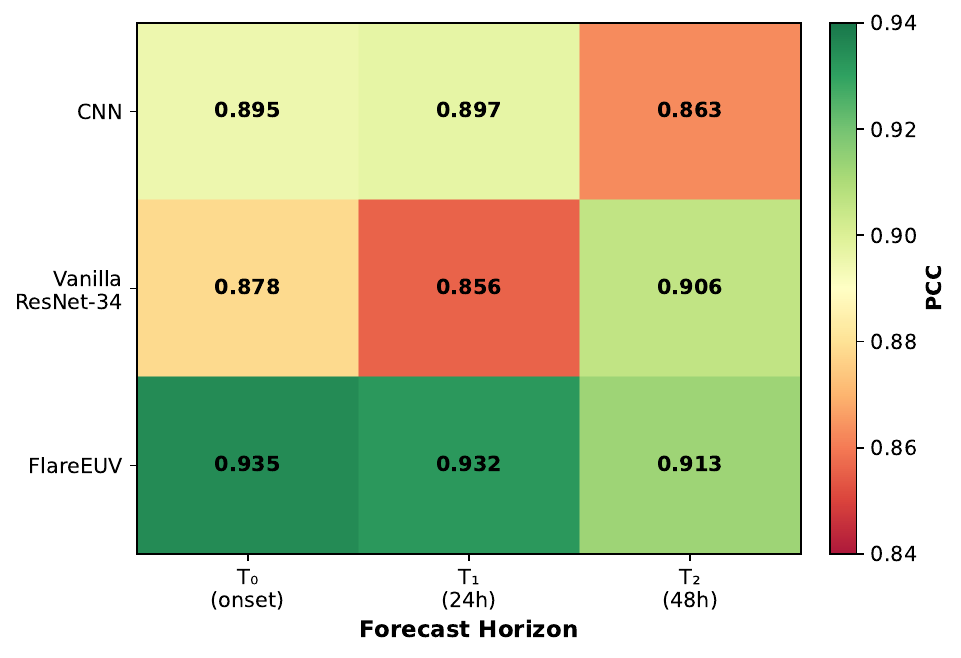}
\caption{Per-horizon Pearson correlation coefficient (PCC) across forecast horizons $T_0$, $T_1$, and $T_2$ for all evaluated models.}
\label{fig:pcc_heatmap}
\end{figure}

\begin{figure}
\centering
\includegraphics[width=\columnwidth]{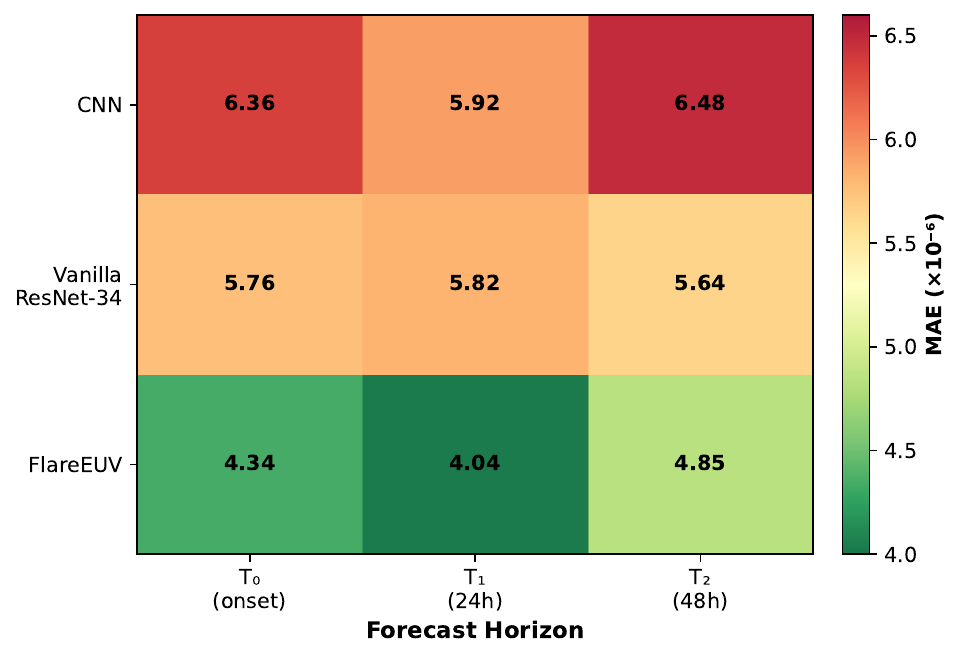}
\caption{Per-horizon mean absolute error (MAE, $\times10^{-6}$ W/m$^{2}$/nm) across forecast horizons $T_0$, $T_1$, and $T_2$ for all evaluated models.}
\label{fig:mae_heatmap}
\end{figure}

Figures~\ref{fig:pcc_heatmap} and~\ref{fig:mae_heatmap}
present per-horizon performance breakdown, revealing that FlareEUV maintains consistent advantages in all three forecast horizons.
The per-horizon analysis reveals two key findings. First, FlareEUV achieves the strongest performance on all horizons, with particularly notable improvements at $T_0$ (flare onset: PCC = 0.935,
MAE = $4.34{\times}10^{-6}$~W/m$^2$/nm) where accurate characterization of peak irradiance is most critical for operational forecasting.  
Second, Vanilla ResNet-34 shows anomalously strong performance at $T_{2}$ (PCC = 0.906) despite weaker near-term forecasting 
(PCC = 0.878 at $T_0$, PCC = 0.856 at $T_{1}$), suggesting that residual connections may preferentially capture long-term decay patterns even without explicit attention mechanisms. However, FlareEUV's consistently superior performance across all horizons (PCC $>$ 0.91 at every horizon) confirms that SE attention provides robust multi-day generalization superior to both lightweight and deep baseline architectures.

\section{Conclusion}
\label{sec:conclusion}
We presented the FlareEUV model, a lightweight 13-channel fusion framework to predict daily
SDO/EVE 6.5\,nm irradiance during significant solar flares in the period 
between 2011 and 2014 in Solar Cycle 24.
By harmonizing AIA radiative observations 
with the HMI magnetic-field structure and employing a horizon-weighted regression objective, the model 
learns stable magnetic–radiative dependencies that generalize across diverse flare events.

Under a rigorous leave-one-flare-out (LOFO) protocol on 33 major flares, FlareEUV achieved a strong global 
correlation of PCC = 0.927 (p $<10^{-40}$) and a mean absolute error of MAE = $4.41 \times 10^{-6}$~W/m$^{2}$/nm, 
demonstrating an accurate prediction of the magnitude of irradiance.
The per-horizon analysis revealed 
consistently strong performance, 
with correlations of 
PCC (at $T_{0}$) = 0.935, 
PCC (at $T_{1}$) = 0.932 and 
PCC (at $T_{2}$) = 0.913, and corresponding mean absolute errors of 
MAE (at $T_{0}$) = $4.34\times10^{-6}$, 
MAE (at $T_{1}$) = $4.04\times10^{-6}$, and 
MAE (at $T_{2}$) = $4.85\times10^{-6}$~W/m$^{2}$/nm.  
The minimal degradation from 
$T_0$ to $T_{2}$ (2.4\% drop in correlation) indicates 
reliable multi-day forecasting capability.
These results substantially outperform both single-modality baselines (AIA-only: PCC = 0.868; 
HMI-only: PCC = 0.835) and architectural baselines (Baseline CNN: PCC = 0.885; 
Vanilla ResNet-34: 
PCC = 0.879), confirming that the combination of multimodal fusion, deep residual architecture, 
and channel attention mechanisms is essential for robust cross-flare generalization. Notably, 
the addition of SE attention blocks adds only 43,520 parameters (+0.2\% overhead) yet yields 
a 5.5\% improvement in correlation over 
Vanilla ResNet-34, demonstrating the efficiency of 
lightweight attention mechanisms for heliophysics applications.

The FlareEUV framework is fully reproducible  
and establishes a compact benchmark 
for multi-instrument irradiance forecasting. 
It should be noted that, 
although FlareEUV was only tested on 
significant flares in the period between 
2011 and
2014
in this study, the framework 
 should generalize to
significant flares in other periods.
We chose this period to demonstrate how FlareEUV works because 2014 is solar maximum in Solar Cycle 24,
with many strong active regions,
significant flares and coronal mass ejections.
In the future, we plan to apply
FlareEUV 
to additional 
solar cycles (e.g., Solar Cycle 25), 
smaller flare classes, and operational real-time forecasting systems to further 
validate cross-cycle generalization and enable deployment in space-weather prediction infrastructure.

\section*{Acknowledgment}

The authors acknowledge NASA/SDO and the EVE, AIA, and HMI instrument teams for providing open access to the data used in this study \cite{lasp_eve_l3,jsoc_aia_data,jsoc_hmi_data}
as well as
the SunPy project community
\cite{sunpy_fido}
for providing 
Python libraries for solar physics 
studies.
This work was supported in part by NASA under grant numbers 80NSSC24M0174 and
80NSSC24K0548.

\bibliographystyle{IEEEtran}

\balance
\end{document}